\documentclass[runningheads]{llncs}

\usepackage{amsmath}
\usepackage[T1]{fontenc}
\usepackage{graphicx}
\usepackage{color}
\usepackage{url}

\usepackage{enumitem}
\usepackage{amsfonts}

\usepackage{tabularx}
\usepackage{booktabs}
\usepackage{multirow}
\usepackage{caption}
\usepackage{subcaption}
\usepackage{cite}

\newcommand{\mypar}[1]{\smallskip\noindent\textbf{#1.}}
\newcommand{\mypartwo}[1]{\vspace{0.5pt}\noindent\textit{#1.}}
\usepackage[hidelinks]{hyperref}

\usepackage{algorithmicx}
\usepackage{algorithm}
\usepackage[noend]{algpseudocode} % 'noend' option removes 'end if', 'end for', etc.
\usepackage{eurosym}
\usepackage{todonotes}

\begin{document}
\title{From Global Policies to Local Strategies: Multi-Objective Optimization of Resource-Specific Handover Policies}
\titlerunning{Multi-Objective Optimization of Resource-Specific Handover Policies}

% \author{Blinded for peer review}
% \institute{}
\author{
Lukas Kirchdorfer \inst{1,2}\orcidID{0000-0003-4713-9328} \and
Artemis Doumeni\inst{3} \and
Han van der Aa\inst{4}\orcidID{0000-0002-4200-4937
}
\and Hugo A. López\inst{3}\orcidID{0000-0001-5162-7936}}
\authorrunning{Kirchdorfer et al.}
\institute{
SAP Signavio, Walldorf, Germany \\ \email{lukas.kirchdorfer@sap.com} \\ \and
Data and Web Science Group, University of Mannheim, Germany \\ \and
DTU Compute, Technical University of Denmark, Kongens Lyngby, Denmark\\
\email{hulo@dtu.dk}\\ \and 
Faculty of Computer Science, University of Vienna, Vienna, Austria
\email{han.van.der.aa@univie.ac.at}}

\maketitle              % typeset the header of the contribution
\begin{abstract}
Efficient resource allocation is a key challenge in business process management, with direct implications for cost, throughput time, and utilization. While recent Reinforcement Learning (RL) approaches have shown promise in deriving adaptive allocation policies, they typically neglect inter-resource collaboration patterns that can strongly influence real-world task handovers. Recognizing this, this paper introduces the first approach for multi-objective optimization of resource-level decision-making, enabling the recommendation of person-specific handover policies. 
To achieve this, our work combines an existing Multi-Agent System-based process simulator with a multi-objective evolutionary algorithm.
The resulting approach produces Pareto-optimal, resource-specific policies that optimize the process across multiple objectives.
Experimental results on synthetic and real-world datasets show that our approach reduces costs by an average of 37\% and waiting time by 58\%, consistently outperforming heuristic baselines and demonstrating the potential of leveraging collaboration-aware optimization to improve process performance.
\end{abstract}
\section{Introduction}

Efficient resource allocation—the task of assigning resources, such as employees, machines, and systems, to process activities—is a central challenge in business process management, as it directly affects key performance indicators such as cost, throughput time, and resource utilization~\cite{zhao2015optimization}. However, finding effective allocations is difficult due to the complexity and interdependence of concurrent process instances, the need for decisions under dynamic conditions, and the long-term consequences of allocation choices~\cite{meneghello2024optimizing}. Accordingly, a wide range of approaches has been proposed, ranging from rule- and scheduling-based methods~\cite{zhao2015optimization} to more recent Reinforcement Learning (RL)-based approaches~\cite{branchRL,meneghello2024optimizing,ZbikowskiOG22}, which showed promising results by balancing short- and long-term objectives.

Despite this progress, existing RL-based approaches exhibit a key limitation: they learn \textit{global} resource-allocation policies based only on (i) which tasks are awaiting execution and (ii) which resources are available to perform them. This allows them to exploit differences in skill, execution time, and cost, but it treats resources as passive recipients of assignments and ignores an important aspect of real organizational behavior: collaboration patterns among resources. In practice, however, resources---especially human resources---are active decision-makers during process execution, influencing not only which work they perform, but also with whom they collaborate. These inter-resource dependencies give rise to characteristic handover and collaboration patterns that shape how work flows through an organization. 
Concretely, we are interested in \emph{handover policies} that specify, for a given resource and completed activity, to which resource the case should be passed next for the subsequent activity.
While accounting for such \textit{local}, resource-specific handovers has been shown to improve the fidelity of as-is process simulations~\cite{Kirchdorfer2025}, existing optimization approaches do not account for these local handovers, and therefore overlook collaboration structure as a leverage for process improvement.

This paper introduces the first approach for grounded multi-objective optimization of resource-specific handover policies in business processes. Our approach takes as input a Multi-Agent System (MAS)-based simulation model of the process that explicitly captures resource interdependencies, building on prior work~\cite{Kirchdorfer2025}. On top of this model, we develop an analysis-and-design optimization framework that integrates the widely used multi-objective evolutionary algorithm NSGA-II~\cite{deb2002} to search for Pareto-optimal handover policies. 
This approach offers two key advantages. First, it can be grounded in data, since the MAS simulation model taken as input can be discovered from event logs (cf.\cite{Kirchdorfer2025}).
Second, it is generic with respect to any set of objectives, yielding a Pareto front of handover policies that offer practitioners actionable alternatives that reflect different operational trade-offs. 
To increase solution diversity, we incorporate four mutation variants into the genetic algorithm. Experiments on synthetic and real-world datasets, using cost and time as example objectives, demonstrate substantial improvements over the as-is process, with average reductions of 37\% in cost and 58\% in waiting time, also outperforming established heuristic baselines.

The remainder of this paper is structured as follows: \autoref{sec:relatedwork} reviews related work, \autoref{sec:problemdefinition} defines the problem, \autoref{sec:approach} presents our approach, and \autoref{sec:evaluation} reports the evaluation. Finally, \autoref{sec:conclusion} concludes the paper.

\section{Related Work}
\label{sec:relatedwork}
Our approach to optimizing resource-specific handover policies relates to three main research directions.

\mypar{Resource allocation}
A large body of work has addressed resource allocation in business processes, mostly focusing on optimizing task--resource assignments during process execution, i.e., at runtime. Early approaches rely on rule-based assignment and scheduling heuristics~\cite{zhao2015optimization}, which depend on predefined rules or short-term objectives and therefore adapt poorly to dynamic process conditions. Prescriptive Process Monitoring (PrPM) approaches~\cite{sindhgatta2016context,wibisono2015fly} extend these methods by recommending resources for ongoing cases, yet they still optimize decisions locally and often ignore their broader process-level impact.

To address longer-term and process-level effects, more advanced work integrates simulation and applies metaheuristic or learning-based optimization methods, such as genetic algorithms~\cite{Djedovic,Lee2001Integration,SI201872}, heuristic search~\cite{OptimusPaper}, or RL~\cite{huang2011reinforcement,meneghello2024optimizing,MiddelhuisBSBAD25,ZbikowskiOG22}. Yet, only a subset considers multi-objective optimization as addressed in this paper. For example, Huang et al.~\cite{huang2011reinforcement} use RL to optimize cost and cycle time.

While our approach can be employed at runtime to determine handover decisions (which we will do in our evaluation), in principle, it optimizes a static model of resource collaboration, independently of runtime case information. Thus, we consider it as design-time.
Within this design-time line of work, several approaches optimize the number of required resources per resource pool~\cite{OptimusPaper,Peters21,Filatov_Yerokhin_2023}, solving a staffing problem rather than optimizing handovers. Closest to our work, Bejarano et al.~\cite{Bejarano23} also address a design-time, simulation-based, multi-objective optimization problem, but their decision variables remain activity-to-resource assignments, whereas ours are handover decisions between resources.

% López-Pintado et al.~\cite{OptimusPaper} generate a Pareto front using hill-climbing to jointly optimize cost and time. However, their approach addresses a different problem: it optimizes the number of resources in a resource pool, effectively solving a staffing problem rather than optimizing collaboration structures.

In summary, a crucial distinction sets our approach apart from all existing work: the underlying representation of resources. Existing methods treat resources as passive entities that are assigned to activities; optimization thus centers on determining which resource should execute which task. In contrast, we adopt a MAS perspective in which resources are first-class citizens, modeled as autonomous decision-making agents capable of choosing their collaboration partners. Consequently, instead of learning a single global allocation rule, our approach discovers \emph{resource-specific handover policies} that capture and optimize the collaborative structure of the process. This enables a form of collaboration-aware optimization that has not been addressed in previous research.

\mypar{Organizational and agent system mining}
Complementary to optimization approaches, organizational and resource-oriented process mining methods analyze how resources collaborate~\cite{SongA08}. More recently, agent system mining has been proposed to discover process control-flow models from an agent perspective~\cite{Tour2021}. However, these approaches remain descriptive: they reveal collaboration structures but do not treat them as decision variables for optimization.

\mypar{Agent-based business process simulation}
Traditional business process simulation typically follows a control-flow-first perspective, extending a process model (e.g., a Petri net) with temporal and resource-related parameters~\cite{ling2000time}. In contrast, agent-based approaches adopt a resource-first perspective in which resources are modeled as autonomous entities at the core of the simulation~\cite{Kirchdorfer2025}. While such agent-based simulations can be used in a \textit{what-if} manner to explore alternative collaboration patterns, this requires manually defining and evaluating different configurations. Our approach automates this process by systematically exploring handover policies and presenting the user with a set of Pareto-optimal collaboration patterns and their corresponding trade-offs.

\section{Problem Definition}
\label{sec:problemdefinition}

\mypar{Event logs} 
Event logs represent a key artifact of our approach, generated during the search for optimal resource policies.
We define an event log $L$ as a collection of traces. A trace $\sigma \in L$ is a finite sequence of events, $\langle e_1,...,e_n\rangle$, recording the execution of activities performed for a single case in a process. Each event $e_i$ is a tuple $(case, act, ts_{start}, ts_{end}, res)$, where $case$ is the case's identifier, $act$ is the activity to which the event corresponds, $ts_{start}$ and $ts_{end}$, respectively, are the start and end timestamps of the activity's execution, and $res$ is the resource that executed the activity. 

\mypar{Pareto optimization}
Our approach aims to improve resource handover policies by optimizing a set of $k > 1$ objective functions. In our setting, an event log $L$ captures the realization of the resource handover policy applied during the execution of its cases. Thus, we define the cost vector of an event log $L$ as
\begin{equation}
\vec{c}(L) = \big[ f_1(L), \dots, f_k(L) \big],
\qquad \text{where } f_i(L) = \sum_{\sigma \in L} f_i(\sigma),\; f_i : L \rightarrow \mathbb{R}.
\label{eq:cost}
\end{equation}

Since our goal is to minimize all components of the cost vector $\vec{c}(L)$, we seek to minimize the corresponding set of objective functions $\mathcal{K} = \{f_1(L),\dots,f_k(L)\}$. As these objectives are frequently in conflict, we focus on identifying solutions that represent optimal trade-offs, formalized through the notion of \emph{Pareto optimality}.
Let $\mathcal{L}_{C}$ denote the set of all event logs defined over a set of cases $C$. For two logs $L, L' \in \mathcal{L}_{C}$, we follow \cite{hugo2024pareto} and say that $L$ \emph{dominates} $L'$ if
\begin{equation}
\forall i \in {1,\ldots,k}: f_i(L) \le f_i(L')
\quad \land \quad
\exists j \in {1,\ldots,k}: f_j(L) < f_j(L').
\label{eq:dominance_log}
\end{equation}

In other words, $L$ performs at least as well as $L'$ on all objectives and strictly better on at least one. A log $L \in \mathcal{L}_{C}$ is \emph{Pareto optimal} if no log $L' \in \mathcal{L}_{C}$ exists that dominates it. The collection of all non-dominated logs constitutes the \emph{Pareto front}, representing the set of optimal trade-offs across the $k$ objectives.

\mypar{The resource allocation problem}
Let $\mathcal{A}$ denote a set of agents, where each agent $a \in \mathcal{A}$ corresponds to a resource $res \in \mathcal{RES}_L$ involved in the process. Likewise, let $\mathcal{ACT}_L$ denote the set of all activities in the process, where $act, act' \in \mathcal{ACT}_L$. 
We denote each tuple $(a, act, act')$ a \emph{decision point} for agent $a$ after completing activity $act$, in the sense that it needs to hand over to some agent $a'$ to work on $act'$.
To model this decision, we define a \emph{handover policy} $\pi$ as a conditional probability distribution:

\begin{equation}
    \pi(a' \mid a, act, act') 
    \qquad \text{where } a, a' \in \mathcal{A},\; \text{and } act, act' \in \mathcal{ACT}_L.
    \label{eq:policy}
\end{equation}

Optimizing $\pi$ thus corresponds to searching for a handover pattern that yields improved process performance with respect to the multi-objective cost vector defined in~\autoref{eq:cost}.

\section{Approach}
\label{sec:approach}

\begin{figure}[t]
    \centering
    \includegraphics[width=\linewidth,
        trim=150 200 10 210,clip]{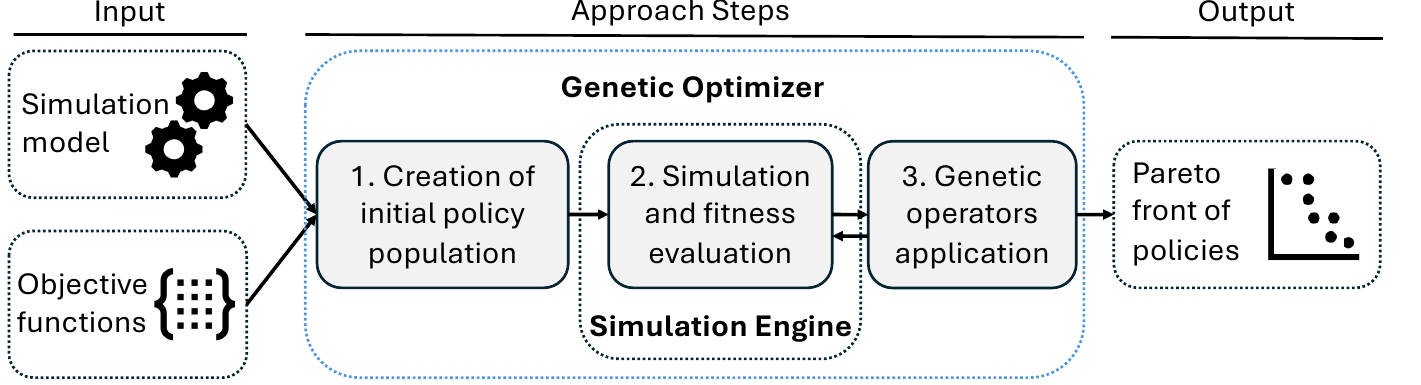}
    \caption{Overview of the proposed approach.}
    \label{fig:approachoverview}
    \vspace{-1.5em}
\end{figure}

This section introduces our genetic algorithm-based approach for discovering Pareto-optimal resource handover policies, as illustrated in \autoref{fig:approachoverview}. Its input, main steps, and output are as follows:

\mypar{Input}
Our approach takes as input i) an agent-based simulation model $\mathcal{M}$ and ii) a set of objective functions $\mathcal{K} = \{f_1(L),\dots,f_k(L)\}$, which instantiate the cost vector (see \autoref{eq:cost}) that guides the multi-objective optimization.
We define the simulation model as
$\mathcal{M} = (\mathcal{A}, \mathcal{S}, \pi_0, \beta)$,
where \(\mathcal{A}\) is the set of agents, \(\mathcal{S}\) captures global simulation parameters, \(\pi_0\) is the as-is handover policy, and \(\beta\) is the control-flow policy that determines the next activity in a running case.
Each agent \(a \in \mathcal{A}\) is characterized by its weekly availability calendar and its capabilities, comprising the set of activities it can perform and the corresponding execution-time distributions. The tuple \(\mathcal{S}\) captures global simulation parameters such as the case interarrival distribution and the arrival calendar.
The simulation model used in our approach can either be specified manually with domain experts or automatically discovered from an event log $L$. In this work, we employ \textit{AgentSimulator}~\cite{Kirchdorfer2025}, which automatically derives a multi-agent simulation model from an event log and supports the simulation of the corresponding process, thereby also serving as the simulation engine. For a detailed description of the quality of the discovered simulation models, we refer to the original publication. Note that other approaches may also be used to instantiate the engine.

\mypar{Approach steps}
The genetic optimizer is the central component of our approach. It takes as input the as-is model $\mathcal{M}$ with its handover policy~$\pi_0$  and optimizes the policy with respect to the specified set of objective functions $\mathcal{K}$. The optimization workflow consists of three steps:

\begin{enumerate}[noitemsep,topsep=0pt]
    \item \textit{Creation of initial policy population.} Based on the as-is policy~$\pi_0$, we generate an initial population of handover policies, denoted by $\mathcal{P} = \{\pi_1, \pi_2, \ldots, \pi_N\}$, consisting of $N$ candidate policies.
    \item \textit{Simulation and fitness evaluation}. For each candidate policy $\pi_n \in \mathcal{P}$ in the population, the optimizer calls the simulation engine to simulate multiple runs.
    These runs yield event logs that are used to compute a multi-objective fitness score for the underlying policy, as defined in \autoref{eq:cost}.
    \item \textit{Genetic operators application.} The optimizer uses the feedback from the fitness evaluation to apply evolutionary operators to its population, generating a new, potentially superior set of candidate policies.
\end{enumerate}

\noindent
We employ NSGA-II~\cite{deb2002}, a genetic algorithm specialized for multi-objective optimization, to guide the evolutionary search. We instantiate NSGA-II to operate on individuals in the population that encode full probabilistic handover policies, and adapt the algorithm through custom genetic operators designed to preserve and meaningfully modify these probability distributions. It is important to note that while the selection of NSGA-II has been driven by speed and usability concerns against alternatives (e.g., NSGA-I~\cite{srinivas1994muiltiobjective}, SPEA~\cite{zitzler1998evolutionary}), our approach is independent of the chosen algorithm.

\mypar{Output}
Our approach repeats steps 2 and 3 a predefined number of iterations, controlled by a hyperparameter $G \in \mathbb{N}$. After $G$ generations, the algorithm yields a Pareto front of handover policies $\mathcal{P}^\star = \{\pi_1, \pi_2, \ldots, \pi_M\}$, that indicate how resources should collaborate to achieve an optimal trade-off.

\noindent In the remainder, we describe the details of its core steps: the creation of an initial population, the simulation and fitness evaluation, and the genetic operators.

\subsection{Creation of the Initial Policy Population}

The first step of the genetic optimizer involves constructing an initial population of handover policies, derived from the as-is policy~$\pi_0$. 
% identified from the input event log. 
In this setting, each policy constitutes an \textit{individual} (also known as the \textit{chromosome}) in the population of the genetic algorithm.

\mypar{Search space pruning}
To reduce computational complexity, we prune the search space through a preprocessing step. We analyze the discovered handover policy and classify each decision point---i.e., each situation where a resource finishes an activity and must select a successor (see definition in \autoref{sec:problemdefinition})---based on the determinism of that decision point:
\begin{enumerate} %[noitemsep,topsep=0pt]
\item \textbf{Fixed:} Decision points with only one successor. These reflect non-negotiable business rules and are excluded from optimization.
\item \textbf{Variable:} Decision points with multiple successors. These represent flexible parts of the process and form the search space for optimization.
\end{enumerate}

\noindent 
Therefore, we reduce the search space while preserving all meaningful degrees of freedom. Note that the fixed part of the policy $\pi_{\text{fix}}$ will be used again during the simulation and fitness-evaluation steps.

\mypar{Initial population}
To balance \textit{exploration} and \textit{exploitation}, we construct an initial population~$\mathcal{P}$ of size~$N$ based on the variable part of the as-is handover policy~$\pi_0$. As a heuristic, half of the individuals are replaced with fully random policies to ensure broad coverage of the search space. The other half is generated by mutating the as-is policy, producing candidates that remain close to the observed process behavior. Finally, one individual is set to an exact copy of the as-is policy to ensure it is included. This hybrid strategy provides both diverse and domain-informed starting points for the evolutionary search.

\subsection{Simulation and Fitness Evaluation}
After generating the initial population of handover policies~$\mathcal{P}$, the next step evaluates the fitness of each individual by means of simulation. 
The fitness evaluation treats the simulator as a black-box function whose outputs are then assessed. As summarized in Algorithm~\ref{alg:fitness_eval}, the \textbf{fitness of a single chromosome} is computed through three stages: policy reconstruction, simulation, and evaluation.

\mypar{Policy reconstruction}
Each $\pi_n \in \mathcal{P}$ represents only the variable part of the handover policy. To obtain a complete policy compatible with the simulator, we merge this variable portion with the fixed part $\pi_{\text{fix}}$ (Line~\ref{line:merge}) identified during the search space pruning. The result is a full, comprehensive handover policy that specifies all resource handover decisions in the process.

\mypar{Simulation}
Using the reconstructed policy $\pi_n \in \mathcal{P}$, we instantiate a new simulation model $\mathcal{M}_t$ (Line~\ref{line:new_model}). Specifically, we begin with the parameters of the initial simulation model $\mathcal{M}$—preserving all settings such as arrival rates and resource schedules—and overwrite only the handover policy parameter (Line~\ref{line:overwrite}). This yields a complete, valid model that can be executed to generate a simulated event log (Line~\ref{line:simulate}).
To mitigate stochastic variance and obtain stable performance estimates, the simulation is run multiple times for each policy, independently generating several event logs. Thus, per policy $\pi_n$, we generate $T$ event logs, denoted by $L_n^1, \dots, L_n^T$.

\mypar{Evaluation}
For each simulated event log $L_n^t$, we compute the cost vector $\mathbf{c}_t$ (Line~\ref{line:compute_metrics}) defined in \autoref{eq:cost}. The final fitness score of the policy $\pi_n$ is obtained by averaging its cost vectors across all simulated logs $L_n^1, \dots, L_n^T$, resulting in a robust multi-objective fitness estimate (Line~\ref{line:compute_fitness}).

\begin{algorithm}[t]
\caption{Simulation and Fitness Evaluation}\label{alg:fitness_eval}
\begin{algorithmic}[1]
\Procedure{EvaluateFitness}{var. policy $\pi_n$, fix. policy $\pi_{\text{fix}}$, $\mathcal{M}$, $T$, $\mathcal{K}$}
    \State $\pi_n \gets \Call{MergePolicyParts}{\pi_n, \pi_{\text{fix}}}$ \Comment{Reconstruct full handover policy} \label{line:merge}
    \State $\mathcal{C} \gets \emptyset$ \Comment{Initialize set of cost vectors for all runs}
    
    \For{$t \gets 1 \textbf{ to } T$}
        \State $\mathcal{M}_t \gets \Call{Copy}{\mathcal{M}}$ \Comment{Instantiate simulation model for run $t$} \label{line:new_model}
        \State $\mathcal{M}_t.\pi \gets \pi_n$ \Comment{Overwrite initial handover policy} \label{line:overwrite}
        
        \State $L_n^t \gets \Call{RunSimulation}{\mathcal{M}_t}$ \Comment{Simulated event log} \label{line:simulate}
        \State $\mathbf{c}_{t} \gets \Call{CalculateMetrics}{L_n^t,\mathcal{K}}$ \Comment{Cost vector for run $t$} \label{line:compute_metrics}
        \State $\mathcal{C} \gets \mathcal{C} \cup \{\mathbf{c}_{t}\}$ \Comment{Append to set of cost vectors}
    \EndFor
    
    \State $\bar{\mathbf{c}} \gets \Call{VectorMean}{\mathcal{C}}$ \Comment{Compute fitness as average cost vector} \label{line:compute_fitness}
    \State \Return $\bar{\mathbf{c}}$
\EndProcedure
\end{algorithmic}
\end{algorithm}

\subsection{Genetic Operators}
Having computed a fitness score for each policy in the population, the next step in our approach is to apply genetic operators to the current population~$\mathcal{P}$. In evolutionary computation, genetic operators are mechanisms that modify or recombine individuals to drive the search process. Inspired by biological evolution, they introduce variation, explore the search space, and steer the population toward higher-quality solutions~\cite{deb2002}. 
In our approach, we employ NSGA-II’s three standard operators—selection, crossover, and mutation—applied sequentially, and customize the latter two to operate on probabilistic handover policies.

\mypar{Selection}
First, the algorithm performs a selection step to retain only the most promising individuals for survival and reproduction through the subsequent crossover and mutation operators. Individuals are ranked into non-dominated fronts based on their fitness scores. Afterwards, pairs of randomly drawn policies are compared, and the one with the superior rank is selected. If both belong to the same front, ties are broken using the \textit{crowding distance}, which measures how isolated a policy is from its neighbors in the objective space, thereby favoring solutions in less-explored regions. This combination of non-domination ranking and crowding-based tie-breaking ensures that NSGA-II selects individuals that promote both convergence toward the Pareto front and diversity along it.

\mypar{Crossover}
Given the selected subset of policies, the crossover operator combines two candidates to produce new policies (\textit{offspring}) that inherit characteristics from both \textit{parents}. In our context, a handover policy consists of a set of decision points, each represented by a probability distribution over possible successor resources. Because mixing individual probabilities can yield invalid distributions, crossover must preserve each decision point as a coherent unit. Our custom operator therefore works at the level of entire decision points: for each decision situation, it randomly swaps the full probability distribution between the two parents. This ensures that the offspring inherit valid decision behaviors while still enabling the exploration of new combinations of strategies within the probabilistic constraints of the handover model. The crossover operation is applied to two parents with probability $p_c$. If there is no crossover, the offspring are exact clones of their parents. This hyperparameter $p_c$ controls the degree of exploitation by determining how frequently the algorithm recombines strategies from high-performing solutions.

\mypar{Mutation}
Mutation serves as the main driver of novelty in our evolutionary search, applied to the previously generated offspring. Whereas crossover recombines existing decision strategies, mutation introduces new behavioral patterns by perturbing selected decision points within a handover policy. Its purpose is to prevent premature convergence and to support continuous exploration of alternative resource allocation behaviors.
In the context of resource handover policies, different business environments may benefit from different mutation dynamics. Some scenarios require strong exploratory jumps to escape local optima, while others benefit more from careful refinements once the population has converged toward a promising region of the search space. To promote robust performance across diverse process structures, we implement several complementary mutation strategies.
Mutation is applied to each offspring with probability $p_m$. If selected for mutation, one of its decision points is chosen at random, and one of four mutation variants is applied to the probability distribution at that point.

\mypartwo{Random} 
The probability distribution at the selected decision point is reinitialized, providing a strong exploratory step by generating a completely new handover strategy for that situation.
Concretely, let the decision point specify a categorical distribution over its admissible successor resources in the variable part of the policy. The operator assigns a new positive weight to each successor, drawn independently from a uniform distribution, and normalizes these weights so that they again sum to one. 
Because this operator does not consult $\mathcal{M}$ or the objective set $\mathcal{K}$, it can realize handover behaviors that are not reachable through small, heuristic-guided shifts. In our implementation, it plays the same role in initializing the population, where fully random variable policies are mixed with mutated variants of $\pi_0$.

\mypartwo{Greedy}
Using a local heuristic derived from $\mathcal{M}$, the operator identifies the best and worst successor resources at the selected decision point and moves all probability mass away from the worst option---an aggressive exploitation step.
Among the objectives in $\mathcal{K}$, the operator first draws one objective uniformly at random. It then scores each successor resource that carries a non-negligible probability at this decision point. The successors with the minimum and maximum scores are retained as best and worst, respectively.
The entire probability allocated to the worst successor is set to zero and transferred to the best successor. If the best successor is associated with several successor activities, the incoming mass is distributed across them in proportion to their current probabilities (or uniformly if it currently receives no mass).

\mypartwo{Guided}
Also relying on the local heuristic above, this variant transfers only a small, random fraction of probability from the worst to the best successor. It is designed for fine-grained adjustments in an already promising region of the search space.
After selecting the same guidable objective from $\mathcal{K}$ and ranking successors as in greedy mutation, guided mutation draws a transfer share uniformly between 10\% and 50\% of the probability mass currently assigned to the worst successor. This amount is subtracted from the worst successor's activities in proportion to their shares and added to the best successor's activities using the same proportional rule as in the greedy operator. The worst successor, therefore, retains most of its former probability, yielding a conservative, heuristic-biased step along one dimension of $\mathcal{K}$ while preserving a valid distribution at the decision point.

\mypartwo{Hybrid}
To balance exploration and exploitation, the operator selects one of the above strategies with predefined probabilities: 60\% guided, 20\% greedy, and 20\% random. This weighting reflects the intuition that, as the search progresses, small refinements are often most effective, while still allowing for larger changes.

Together, these mutation strategies enable the genetic algorithm to flexibly adapt its search behavior, supporting both incremental improvements and substantial shifts in handover logic when needed. While we employ these four mutations, other heuristics can also be incorporated. 

Thus, the genetic operators generate a new population of handover policies, which is then again used for simulation and fitness evaluation. We repeat this evolutionary cycle for a predefined number of $G$ iterations, ultimately yielding a Pareto front of resource handover policies $\mathcal{P}^\star$.

\section{Evaluation}
\label{sec:evaluation}

This section presents the experiments used to evaluate the performance of our multi-objective optimization approach. We instantiate the set of objective functions with \textit{waiting time} and \textit{labor cost}, thus assessing whether we can learn resource-specific policies that improve both cost and time compared to the as-is scenario and other baselines. For internal validity, we assess the effectiveness of our approach in a controlled setting designed to test its performance across diverse scenarios. For external validity, we evaluate the effectiveness of our approach on seven widely used event logs from business process simulation research, providing an ideal competitive benchmark that lets us assess whether our approach can learn more optimal policies for these processes.
In the remainder, \autoref{sec:setup} describes the experimental setup, followed by the results in \autoref{sec:results}. Implementations and additional results can be found in our repository\footnote{\url{https://github.com/lukaskirchdorfer/BPS-MAS-Handover-Optimizer}}.
% \footnote{\url{https://github.com/lukaskirchdorfer/AgentSimulator-Optimizer}}.

\subsection{Experimental Setup}
\label{sec:setup}
The experimental setup is identical across both experiments, differing only in the datasets used.

\mypar{Internal validity}
To investigate the performance of our approach, we study 5 variants of an existing synthetic \textit{Loan Application} process~\cite{chapela2025}, allowing us  to assess the capability of our approach to learn effective policies across varying process configurations. 
% For the first part of our evaluation, we use an existing synthetic \textit{Loan Application} process as the basis (cf.~\cite{chapela2025}; see~\autoref{tab:datasets} for details) and derive multiple variants to systematically assess the capability of our approach to learn effective policies across varying process configurations.
The process consists of 12 activities, including parallelism, choices, and a loop.
% starting with \emph{Check application form completeness}. Its control-flow structure includes one loop, a parallel branch comprising three activities, three exclusive gateways, and three possible end events: \emph{Approve application}, \emph{Reject application}, and \emph{Cancel application}. 
By default, the process is executed by 19 distinct resources, divided into 6 roles. To create the variants, we employ \textit{AgentSimulator} \cite{Kirchdorfer2025} to discover a simulation model of the as-is process from the original event log, which we then modify. For each variation of the simulation model, we generate an event log containing 1000 cases, consistent with the size of the original log. Specifically, we create the following variants:
\begin{itemize}[noitemsep, topsep=0pt]
    \item Loan$_{AD}$: Cost and duration of activities are equal for all resources.
    \item Loan$_{JS}$: For each activity, resources are split into junior and senior employees, where juniors are cheaper but require more time to execute the task.
    \item Loan$_{RC}$: Creating resource contention by halving the available resources.
    \item Loan$_{SCDT}$: All resources cost the same, but differ in execution times.
    \item Loan$_{ARR}$: Increasing workload through a higher rate of case arrivals.
\end{itemize}

\mypar{External validity}
To study how our approach could optimize logs in the process mining community, we employ seven widely used event logs (see \autoref{tab:datasets}) from business process simulation research \cite{Kirchdorfer2025,Lopez-PintadoD22} as a basis for optimization scenarios. They are well-suited for our analysis as they provide both start and end timestamps for each event, which is required for simulation. The datasets span multiple domains, including financial services, education, and procurement, and differ in size and complexity (datasets available in our repository). 

% \vspace{-2em}
\begin{table*}[t]
\centering
\setlength\tabcolsep{5pt} % default value: 6pt
\caption{Description of event log properties.}
\label{tab:datasets}
\begin{tabular}{lrrrrrr}
%\hline
\toprule
\textbf{Log} & \textbf{Type} & \textbf{Traces} & \textbf{Events} & \textbf{Activities} & \textbf{Agents} & \textbf{Avg. cycle time} \\ 
\midrule
 Loan App & syn & 1000 & 7492 & 12 & 19 & 10  \\ 
 P2P & syn & 608 & 9119 & 21 & 27 &  515 \\ 
 C1000 & syn & 1000 & 38160 & 42 & 14 & 22 \\ 
 C2000 & syn & 2000 & 77418 & 42 & 14 & 20  \\ 
 ACR & real & 954 & 6870 & 18 & 432 & 357 \\  
 BPI12W & real & 8616 & 59302 & 6 & 52 & 214  \\ 
 BPI17W & real & 30276 & 240854 & 8 & 136 & 304  \\ 

 \hline
\end{tabular}
\vspace{-1em}
\end{table*}
% \vspace{-2em}

A limiting factor for external validity is that existing logs do not include information on resource cost allocation. This information is essential for our multi-objective analysis, for which we decided to provide an approximation. We define hourly cost values for each resource. While assigning uniform costs to all resources provides a simple baseline, it prevents the exploration of cost-time trade-offs. 
Therefore, we derived a 
% To create a more plausible optimization scenario, a synthetic, 
a performance-correlated cost model for each event log. This reflects the real-world principle that more efficient or specialized resources are typically more expensive. To achieve this, we created a performance profile for each resource based on skills and activity durations. We then clustered the profiles into five distinct tiers using k-means, assigned an hourly cost range to each tier with higher-performing tiers associated with higher costs (e.g., Tier 1: 10-25\$/hr, \dots, Tier 5: 76-90\$/hr). We randomly sampled a value for each resource from its corresponding cost range.

\mypar{Benchmark approaches}
% We benchmark our resource allocation approach against several baselines:
We compare against several baselines\footnote{While we initially wanted to include \cite{huang2011reinforcement} in our evaluation, their method employs tabular Q-learning, which requires storing all state–action values and thus does not scale to realistic BPM state spaces. In addition, the approach lacks a reproducible implementation and was evaluated only in a custom simulator. We therefore exclude it from the empirical comparison.}:

\begin{itemize}[noitemsep, topsep=0pt]
    \item \textbf{As-Is}: The as-is resource handover patterns discovered by AgentSimulator.
    \item \textbf{Avail}: Always hand over to the earliest available resource.
    \item \textbf{Random}: Select a random resource.
    \item \textbf{LC}: Always choose the cheapest resource.
    \item \textbf{SPT}: Select the resource with the shortest expected processing time.
\end{itemize}

\mypar{Hyperparameters}
We use the following hyperparameter configuration for all experiments. We use a population size $N=100$ and run $G=100$ iterations. To account for simulation stochasticity, each policy is evaluated using $T=3$ simulated logs. The probability that crossover is applied is set to $p_c = 70\%$. The probability that a mutation is applied to an offspring is set to $p_m = 30\%$. As detailed in \autoref{sec:approach}, we employ four mutation variants for our genetic operator (\textit{Random, Greedy, Guided, Hybrid}), all of which we evaluate in our experiments.

\mypar{Metrics}
We use two types of metrics. The first assesses overall process performance under different handover policies by measuring total process cost in \$ (sum of event durations in hours multiplied by the resource’s hourly rates) and waiting time (WT), defined as the sum of idle times.
The second type assesses the quality of the discovered Pareto fronts, evaluating how effectively each mutation variant explores a diverse and high-quality set of optimal solutions, following the methodology of López-Pintado et al. \cite{OptimusPaper}. These metrics include the Hyperarea Ratio, Purity, Averaged Hausdorff Distance, and Delta Spread. As this analysis is not our primary focus, we summarize the key findings here, while detailed metric definitions and results are available in our repository.

\subsection{Results}
\label{sec:results}
% We begin by discussing the results of our first evaluation, which uses the different modifications of the \emph{Loan Application} process, before detailing the results of the second evaluation.

\mypar{Internal validity}
\autoref{tab:synthetic} shows the performance of the different approaches in terms of WT and cost across the variants of the \emph{Loan Application} process. Unlike the baseline methods, our approach produces an entire set of trade-off solutions. For comparability, we report the performance of the \emph{most balanced solution} from the discovered Pareto front, defined as the solution with the smallest Euclidean distance to the ideal point in the objective space (i.e., zero cost and zero WT).

\begin{table}[t]
\centering
\setlength\tabcolsep{3pt} % default value: 6pt
\caption{Results for the \textit{Loan Application} variants. The best results per variant are highlighted in bold. Cost measured in thousand \$, WT in hours.}
\label{tab:synthetic}
\begin{tabular}{l *{6}{rr}}
\toprule
& \multicolumn{2}{c}{As-Is} & \multicolumn{2}{c}{Avail} & \multicolumn{2}{c}{Random} & \multicolumn{2}{c}{LC} & \multicolumn{2}{c}{SPT} & \multicolumn{2}{c}{Ours} \\
\cmidrule(lr){2-3} \cmidrule(lr){4-5} \cmidrule(lr){6-7} \cmidrule(lr){8-9} \cmidrule(lr){10-11} \cmidrule(lr){12-13} 
 & Cost  & WT & Cost  & WT & Cost  & WT & Cost  & WT & Cost  & WT & Cost  & WT \\

\midrule
Loan$_{AD}$ & 213 & 16.81 & 161 & \textbf{0.43} & 238 & 1.18 & 145 & 6.22 & 164 & 23.93 & \textbf{120} & 0.52 \\
Loan$_{JS}$ & 246 & 1.46 & 231 & 0.65 & 333 & 1.39 & 194 & 26.79 & 212 & 1.00 & \textbf{166} & \textbf{0.59} \\
Loan$_{RC}$ & 67 & 2.30 & 74 & 2.20 & 76 & 2.10 & 59 & 4.14 & 85 & 2.57 & \textbf{49} & \textbf{2.03} \\
Loan$_{SCDT}$ & 254 & 2.78 & 209 & 0.71 & 287 & 1.27 & 204 & 0.65 & 237 & 26.85 & \textbf{179} & \textbf{0.60} \\
Loan$_{ARR}$ & 88 & 0.74 & 83 & \textbf{0.59} & 81 & 0.73 & 57 & 3.84 & 97 & 1.38 & \textbf{53} & 0.69 \\

% \midrule
% Average & \\

\bottomrule
\end{tabular}
\end{table}

Overall, our approach consistently outperforms the as-is baseline in both cost and time across all five scenarios. For example, in the Loan$_{SCDT}$ scenario---where resources share identical hourly costs but differ in execution times---our approach reduces the total cost from \$254k to \$179k and the WT from 2.78 to 0.60 hours. This demonstrates that our approach effectively enhances individual handover policies by prioritizing faster resources while still engaging slower ones to prevent queues and excessive waiting times.

Compared to the other baselines, our approach achieves the lowest cost in all scenarios and the lowest waiting time in three scenarios, while ranking second-best in the remaining two. These results demonstrate that we learn more effective resource handover policies than those currently implemented. Notably, it also surpasses greedy heuristics such as \emph{LC}, which always selects the cheapest resource, by accounting for the long-term consequences of assignment decisions and jointly optimizing cost and temporal performance.

\begin{figure}[t]
    \centering
    \includegraphics[trim = 0mm 0mm 0mm 0mm, clip, width=1.0\columnwidth]{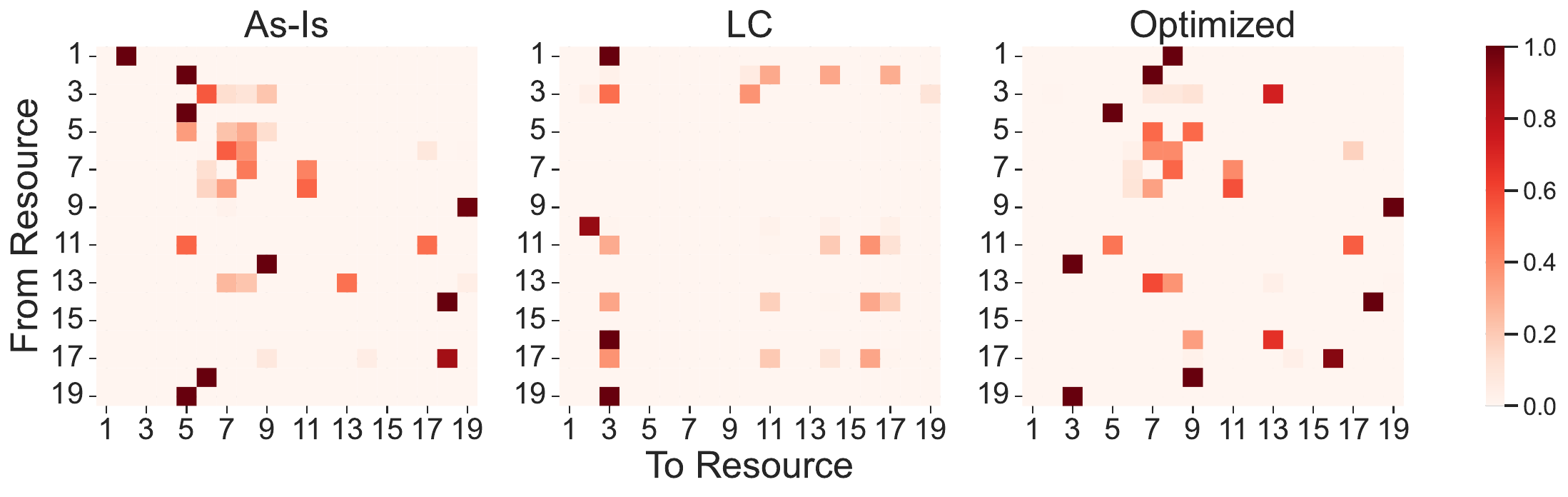}
    \caption{Resource handover frequencies in the Loan$_{JS}$ log based on the \textit{As-Is}, \textit{LC}, and our optimized policy. Darker cells indicate more frequent handovers.}
    \label{fig:loan_js}
    \vspace{-1.5em}
\end{figure}

To better understand these effects, we analyze the resulting handover patterns for the Loan$_{JS}$ log in \autoref{fig:loan_js}, which captures how often resources hand over work to others in the simulated event logs based on the \emph{As-Is}, \emph{LC}, and our optimized policy. The differences between the three are striking. In the \emph{As-Is} baseline, for example, resource~1 always hands over to resource~2. In the \emph{LC} baseline, which optimizes purely for cost, resource~1 instead always hands over to resource~3---who performs the same role but at a lower hourly rate. However, as all resources follow this greedy strategy, costly but faster resources are excluded entirely, leading to severe queue formation, as evident from the many blank cells in \autoref{fig:loan_js}.
Our approach, in contrast, learns a more balanced policy. Again, looking at resource~1, it now primarily hands over to resource~8---a senior costing \$150/hr compared to resource~2’s \$50/hr, but completing tasks six times faster. This makes resource~8 both quicker and more cost-efficient overall. Moreover, unlike the \emph{LC} baseline, where work accumulates at cheap but slow resources, our approach learns individual policies for each resource that, through multi-agent interactions, lead to more complementary collaboration patterns and a more balanced task distribution, resulting in a globally superior outcome.

\mypar{External validity}
The results across the seven existing event logs are presented in \autoref{tab:real}. Overall, consistent with the findings from the first experiment, our approach achieves the best overall performance across most event logs. Importantly, it improves cost and waiting time compared to the \textit{As-Is} baseline in all datasets, underscoring the robustness of our approach in balancing efficiency and cost-effectiveness across diverse real-world settings. 
% (i.e., our approach can learn resource-specific handover policies that enhance the overall process performance). 
While our approach yields particularly robust results in terms of WT, it does not always attain the lowest cost, ranking above the \emph{LC} baseline in four of the seven datasets. 
Moreover, these gains come at the cost of higher computation time: whereas the baselines require about three minutes on average, our approach is roughly 35 times slower (cf. \autoref{tab:runtime}). Like the baselines, runtime increases with the number of events and agents in the log. At the same time, the optimization component scales comparably: for example, BPI17W takes 39 times longer than LoanApp for the baselines and 41 times longer for our approach.
% Moreover, these improvements come at the cost of increased computation: the baselines require, on average across all logs, about three minutes to run, while our approach adds roughly a factor of 35 to this runtime (see details in our repository).

\begin{table}[t]
\centering
\setlength\tabcolsep{3pt} % default value: 6pt
\caption{Results over the existing logs. The best results per log are highlighted in bold. Cost measured in thousand \$, WT in hours.}
\label{tab:real}
\begin{tabular}{l *{6}{rr}}
\toprule
& \multicolumn{2}{c}{As-Is} & \multicolumn{2}{c}{Avail} & \multicolumn{2}{c}{Random} & \multicolumn{2}{c}{LC} & \multicolumn{2}{c}{SPT} & \multicolumn{2}{c}{Ours} \\
\cmidrule(lr){2-3} \cmidrule(lr){4-5} \cmidrule(lr){6-7} \cmidrule(lr){8-9} \cmidrule(lr){10-11} \cmidrule(lr){12-13} 
 & Cost  & WT & Cost  & WT & Cost  & WT & Cost  & WT & Cost  & WT & Cost  & WT \\

\midrule
LoanApp & 91 & 0.71 & 91 & 0.64 & 90 & 0.70 & \textbf{62} & 2.70 & 88 & 0.85 & 70 & \textbf{0.61} \\
P2P & 98 & 17.10 & 116 & \textbf{0.79} & 144 & 2.91 & 61 & 15.26 & 103 & 2.45 & \textbf{44} & 3.62 \\
C1000 & 160 & 1.68 & 133 & 1.58 & 170 & 1.81 & 120 & 1.53 & 121 & 1.49 & \textbf{98} & \textbf{0.75} \\
C2000 & 316 & 1.05 & 251 & 1.03 & 297 & 1.05 & 247 & 1.22 & 237 & 1.03 & \textbf{211} & \textbf{0.71} \\
ACR & 103 & 8.97 & 140 & 0.89 & 232 & 1.01 & \textbf{23} & 3.50 & 96 & 1.64 & 45 & \textbf{0.51} \\
BPI12W & 244 & 0.03 & 177 & \textbf{0.02} & 265 & 0.20 & \textbf{24} & 0.04 & 113 & 0.07 & 176 & \textbf{0.02} \\
BPI17W & 654 & 0.38 & 714 & 0.02 & 675 & 0.16 & \textbf{209} & 2.61 & 317 & 0.43 & 480 & \textbf{0.01} \\

% \midrule
% Average & \\

\bottomrule
\end{tabular}
\end{table}

We also used these event logs to investigate the performance of the four different mutation variants of our genetic operator in \autoref{tab:mutations}, reporting the most balanced solution per variant. This comparison reveals that the \emph{Guided} variant achieves the greatest average improvement across all datasets for both cost and WT relative to the \textit{As-Is} baseline. However, the performance gap between variants is small, indicating that the approach is stable with respect to the mutation strategy.
An analysis of the quality of the resulting Pareto fronts reveals that, while no single variant consistently dominates, the \emph{Guided} and \emph{Hybrid} variants stand out, whereas \textit{Random} and \textit{Greedy} tend to produce less dominant solutions. 
% (further details in our repository).

\begin{table}[t]
\centering
\setlength\tabcolsep{4pt} % default value: 6pt
\caption{Relative improvement in cost and WT per mutation variant compared to the As-Is baseline, averaged across the seven existing logs.}
\label{tab:mutations}
\begin{tabular}{*{4}{cc}}
\toprule
 \multicolumn{2}{c}{Random} & \multicolumn{2}{c}{Greedy} & \multicolumn{2}{c}{Guided} & \multicolumn{2}{c}{Hybrid}  \\
\cmidrule(lr){1-2} \cmidrule(lr){3-4} \cmidrule(lr){5-6} \cmidrule(lr){7-8} 
  Cost  & WT & Cost  & WT & Cost  & WT & Cost  & WT \\
% \midrule
% LoanApp & 22.10\% & 18.61\% & 22.82\% & 13.79\% & 20.19\% & 18.26\% & 22.24\% & 6.26\% \\
% P2P & 28.93\% & 80.16\% & 31.42\% & 77.03\% & 54.74\% & 78.86\% & 32.30\% & 84.21\% \\
% C1000 & 35.88\% & 51.83\% & 38.88\% & 55.35\% & 33.67\% & 52.36\% & 40.32\% & 40.68\% \\
% C2000 & 30.55\% & 34.68\% & 32.39\% & 28.39\% & 34.67\% & 26.00\% & 33.14\% & 32.32\% \\
% ACR & 68.04\% & 79.14\% & 51.92\% & 85.81\% & 62.40\% & 90.60\% & 55.00\% & 88.61\% \\
% BPI12W & 25.38\% & 0.75\% & 27.40\% & 23.75\% & 27.80\% & 41.38\% & 21.99\% & 35.75\% \\
% BPI17W & 23.32\% & 96.30\% & 25.04\% & 97.81\% & 24.71\% & 96.54\% & 26.53\% & 97.69\% \\
\midrule
 33.46\% & 51.64\% & 32.84\% & 54.56\% & \textbf{36.88\%} & \textbf{57.71\%} & 33.07\% & 55.07\% \\
\bottomrule
\end{tabular}
\vspace{-2em}
\end{table}

\begin{table}[htbp]\vspace{-1cm}
\centering
\setlength\tabcolsep{2pt} % default value: 6pt
\caption{Runtimes for all 7 logs (in minutes).}
\label{tab:runtime}
\begin{tabular}{l|ccccccc|c}
\toprule
 & LoanApp & 
 P2P & 
 C1000 & 
 C2000 & 
 ACR & 
 BPI12W & 
 BPI17W & 
 Average \\
\midrule
Baselines 
& 0.28
& 0.35
& 1.78
& 3.58
& 1.72
& 2.17
& 10.83
& 2.96
\\
Ours 
& 9.33
& 16.03 
& 70.67
& 142.60
& 41.83 
& 67.31
& 379.17
& 103.85
\\
\bottomrule
\end{tabular}
% \vspace{-1em}
\end{table}

% To further assess the quality of the solutions discovered by each variant, we analyze the corresponding Pareto fronts in \autoref{tab:pf_quality_transposed}. While no single variant consistently dominates across all four metrics, the \emph{Guided} and \emph{Hybrid} approaches stand out. The \emph{Guided} variant performs best in terms of \emph{Purity}, demonstrating its effectiveness in identifying dominant solutions—aligning with its superior performance in \autoref{tab:mutations}—whereas the \emph{Hybrid} variant achieves the lowest \emph{AHD} values, indicating high convergence toward the reference front. Both variants also achieve similar \emph{HR} values, often close to 1, meaning that their Pareto fronts cover nearly the same area in the objective space as the global reference front. The \emph{Greedy} variant performs best on the \emph{Delta Spread} metric, suggesting a slightly more uniform distribution of solutions, while the \emph{Random} variant consistently produces the lowest-quality fronts across all measures.

\begin{figure}[t]
% \vspace{-2em}
    \centering
    \includegraphics[trim = 0mm 0mm 0mm 0mm, clip, width=0.8\columnwidth]{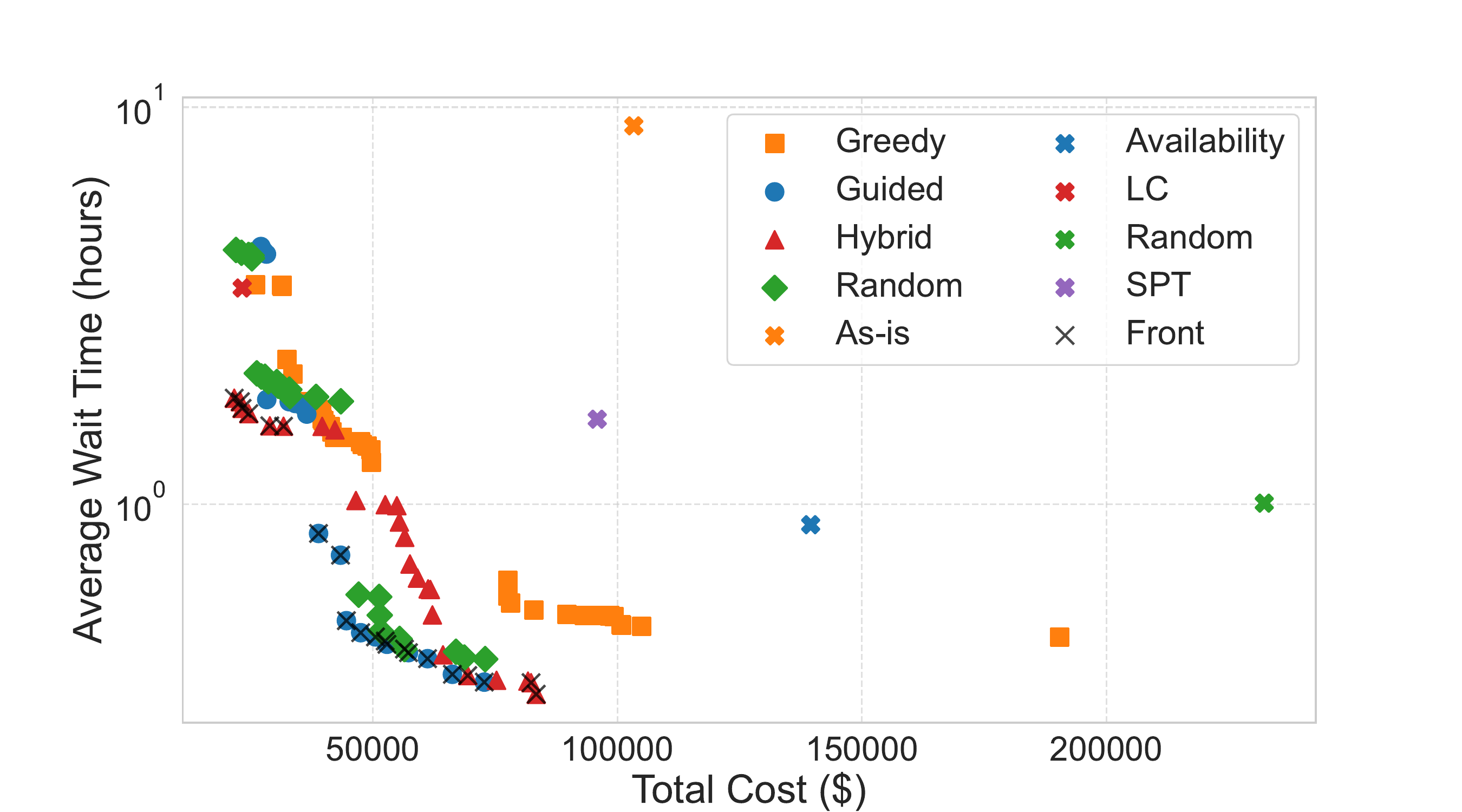}
    \caption{Solution space and Pareto front for ACR event log.}
    \label{fig:ACR}
    % \vspace{-2em}
\end{figure}

% We further illustrate the solution space and Pareto front for the real-life ACR event log in \autoref{fig:ACR}, comparing the set of solutions produced by our approach with the baselines. The \emph{Guided} and \emph{Hybrid} variants clearly dominate the Pareto front, revealing a wide range of Pareto-optimal trade-offs. For example, the total process cost varies between \$22k and \$85k, while the average waiting time ranges from 0.2 to nearly 2 hours.
We further illustrate the solution space and Pareto front for the real-life ACR event log in \autoref{fig:ACR}, comparing our solutions with the baselines. The \emph{Guided} and \emph{Hybrid} variants clearly dominate, offering Pareto-optimal trade-offs with process costs between \$22k and \$85k and waiting times from 0.2 to nearly 2 hours.
While the most balanced solution reported in \autoref{tab:real} corresponds to \$45k and 0.51 hours, our approach offers a diverse set of high-quality alternatives, enabling practitioners and process owners to select policies aligned with their specific operational goals. For instance, if minimizing waiting time is prioritized over cost, one could choose a policy achieving around 0.2 hours, representing an improvement of 0.3 hours over the most balanced solution, 0.7 hours over the heuristic policy that always selects the next available resource, and a remarkable 8.7 hours improvement compared to the as-is status. Likewise, the Pareto front also includes solutions that outperform the \textit{LC} heuristic in terms of cost, showing that our approach can offer both broad coverage and best-in-class solutions.

\section{Discussion and Conclusion}
\label{sec:conclusion}

In this work, we presented a novel approach for optimizing resource-specific handover policies in business processes, enabling the discovery of more effective collaboration strategies. Building on a multi-agent business process simulation model where resources and their interactions are represented as autonomous agents, we integrated a genetic algorithm to optimize process performance across multiple dimensions. The approach is designed to be fully generic: it can incorporate any set of optimization objectives. Its output is a Pareto front of optimized handover policies, providing decision-makers with a transparent set of trade-offs. We validated the approach on synthetic and real logs with cost and time as objectives, achieving 
substantial improvements
% average reductions of 37\% in cost and 58\% in waiting time 
compared to the as-is process.

\mypar{Threats to validity}
Our approach relies on a simulation model to evaluate candidate handover policies, which introduces several validity threats. First, optimization outcomes depend on simulation fidelity: if the simulator does not accurately reproduce real process dynamics, estimated improvements may not fully transfer to reality. To mitigate this risk, we deliberately use a data-driven simulation approach that has been shown in prior work to achieve state-of-the-art accuracy~\cite{Kirchdorfer2025}. Moreover, the simulation model itself is not part of our optimization method but serves as an interchangeable input artifact; therefore, improvements in simulation fidelity can directly benefit optimization outcomes without changes to the proposed approach. Nevertheless, future work should explicitly quantify simulator accuracy and analyze the sensitivity of optimization results to modeling errors.
Second, policies are optimized and evaluated within the same model, which creates a risk of overfitting to simulator-specific dynamics. In practice, this could lead to policies that perform well in simulation but less so in reality. We partially mitigate this threat by grounding simulations in discovered models derived from real event logs and by validating the approach on multiple datasets. A promising direction for future research is robustness analysis across alternative simulation models or perturbed model parameters to assess the stability of discovered policies.
Third, our experiments augment event logs with hourly resource cost information that is typically not recorded in standard logs. While such augmentation is common in process optimization studies, it introduces assumptions that may affect absolute performance estimates. We partially mitigate this threat by using realistic cost assumptions that correlate with performance and by relying on the same assumptions for as-is and optimized policies. Future work could incorporate empirically observed cost data or uncertainty-aware optimization to reduce this dependency further.
Furthermore, our approach assumes that resources can choose whom to collaborate with, or at least that handovers between resources are subject to change. While this assumption holds for many real-world processes---particularly knowledge-intensive processes in which resources have some degree of decision-making autonomy---it may be less applicable to highly orchestrated workflow processes. We refer  to~\cite{Kirchdorfer2025} for details on the types of processes benefiting from resource-specific modeling.

\mypar{Future work}
Beyond validity considerations, the approach entails higher computational effort due to evolutionary search. Although NSGA-II provides a strong baseline for multi-objective optimization, other algorithms may be preferable as the number of objectives grows. Reducing computational overhead---e.g., by narrowing the search space, identifying the most influential decision points, or optimizing only selected handovers---remains an important direction. Furthermore, extending the policy space towards richer collaboration strategies (e.g., dynamic parallelization or adaptive routing) is a promising avenue for future work.
Finally, our evaluation relies on heuristic baselines for comparison. This is partly due to the lack of suitable existing benchmarks addressing the same problem of resource handover optimization. Future work may add stronger optimization-oriented resource allocation baselines that optimize a single objective but add further insights into the value of our approach.

\paragraph{Acknowledgments}
This work received support from VILLUM FONDEN (grant VIL57420) through the Center for Digital Compliance (DICE), the Innovation Fund Denmark projects ``Explainable Hybrid-AI for Computational Law and Accurate Legal Chatbots'' (XHAILe, grant 4355-00018B) and  Predictive and Prescriptive Process Analytics for Industry 4.0 (P3AI4, grant 4105-00045B ).

\bibliographystyle{splncs04}
\bibliography{bibliography}

@inproceedings{ling2000time,
  title={Time Petri nets for workflow modelling and analysis},
  author={Ling, Sea and Schmidt, Heinz},
  booktitle={SMC},
  volume={4},
  pages={3039--3044},
  year={2000},
  organization={IEEE}
}

@article{OptimusPaper,
  author    = {L{\'{o}}pez-Pintado, Orlenys and Dumas, Marlon and Berx, Jonas},
  title     = {Discovery, simulation, and optimization of business processes with differentiated resources},
  journal   = {Inf. Syst.},
  year      = {2024},
  volume    = {120},
  pages     = {102289},
}

@inproceedings{Lopez-PintadoD22,
  author       = {Orlenys L{\'{o}}pez{-}Pintado and
                  Marlon Dumas},
  title        = {Business Process Simulation with Differentiated Resources: Does it
                  Make a Difference?},
  booktitle    = {BPM},
  publisher    = {Springer},
  year         = {2022},
  bibsource    = {dblp computer science bibliography, https://dblp.org}
}

@inproceedings{zhao2015optimization,
  title={The optimization of resource allocation based on process mining},
  author={Zhao, Weidong and Yang, Liu and Liu, Haitao and Wu, Ran},
  booktitle={ICIC},
  pages={341--353},
  year={2015},
}

@inproceedings{sindhgatta2016context,
  title={Context-aware analysis of past process executions to aid resource allocation decisions},
  author={Sindhgatta, Renuka and Ghose, Aditya and Dam, Hoa Khanh},
  booktitle={CAISE},
  pages={575--589},
  year={2016},
}

@inproceedings{wibisono2015fly,
  title={On-the-fly performance-aware human resource allocation in the business process management systems environment using Na{\"\i}ve Bayes},
  author={Wibisono, Arif and Nisafani, Amna Shifia and Bae, Hyerim and Park, You-Jin},
  booktitle={AP-BPM},
  pages={70--80},
  year={2015},
  organization={Springer}
}

@article{deb2002,
  title     = {A fast and elitist multiobjective genetic algorithm: NSGA-II},
  author    = {Deb, Kalyanmoy and Pratap, Amrit and Agarwal, Sameer and Meyarivan, T.},
  journal   = {IEEE TEVC},
  volume    = {6},
  number    = {2},
  pages     = {182--197},
  year      = {2002},
  publisher = {IEEE}
}

@article{srinivas1994muiltiobjective,
  title={Muiltiobjective optimization using nondominated sorting in genetic algorithms},
  author={Srinivas, Nidamarthi and Deb, Kalyanmoy},
  journal={Evolutionary computation},
  volume={2},
  number={3},
  pages={221--248},
  year={1994},
  publisher={MIT Press}
}

@inproceedings{hugo2024pareto,
  author    = {Diaz, Juan F. and L{\'o}pez, Hugo A. and Quesada, Luis and Rosero, Juan C.},
  title     = {Pareto-Optimal Trace Generation from Declarative Process Models},
  booktitle = {BPM Workshops},
  year      = {2024},
  pages     = {314--325}
}

@inproceedings{meneghello2024optimizing,
  title={Optimizing resource allocation policies in real-world business processes using hybrid process simulation and deep reinforcement learning},
  author={Meneghello, Francesca and Middelhuis, Jeroen and Genga, Laura and Bukhsh, Zaharah and Ronzani, Massimiliano and Di Francescomarino, Chiara and Ghidini, Chiara and Dijkman, Remco},
  booktitle={BPM},
  pages={167--184},
  year={2024},
  organization={Springer}
}

@article{huang2011reinforcement,
  title={Reinforcement learning based resource allocation in business process management},
  author={Huang, Zhengxing and van der Aalst, Wil MP and Lu, Xudong and Duan, Huilong},
  journal={DKE},
  volume={70},
  number={1},
  pages={127--145},
  year={2011},
  publisher={Elsevier}
}

@article{zitzler1998evolutionary,
  title={An evolutionary algorithm for multiobjective optimization: The strength pareto approach},
  author={Zitzler, Eckart and Thiele, Lothar},
  journal={TIK report},
  volume={43},
  year={1998},
  publisher={ETH Zurich}
}

@inproceedings{Tour2021,
  author    = {Tour, Andrei and Polyvyanyy, Artem and Kalenkova, Anna},
  title     = {Agent System Mining: Vision, Benefits, and Challenges},
  booktitle = {IEEE Access},
  year      = {2021},
  volume    = {9},
  pages     = {98336-98350},
}

@article{branchRL,
  title={Recommending the optimal policy by learning to act from temporal data},
  author={Branchi, Stefano and Buliga, Andrei and Di Francescomarino, Chiara and Ghidini, Chiara and Meneghello, Francesca and Ronzani, Massimiliano},
  journal={arXiv preprint arXiv:2303.09209},
  year={2023}
}

@article{chapela2025,
title = {A framework for measuring the quality of business process simulation models},
journal = {Information Systems},
volume = {127},
pages = {102447},
year = {2025},
issn = {0306-4379},
_doi = {https://doi.org/10.1016/j.is.2024.102447},
_url = {https://www.sciencedirect.com/science/article/pii/S0306437924001054},
author = {David Chapela-Campa and Ismail Benchekroun and Opher Baron and Marlon Dumas and Dmitry Krass and Arik Senderovich},
keywords = {Business process simulation, Process mining, Quality measures},
}

@INPROCEEDINGS{Djedovic,
  author={Djedović, Almir and Žunić, Emir and Avdagić, Zikrija and Karabegović, Almir},
  booktitle={BIHTEL}, 
  title={Optimization of business processes by automatic reallocation of resources using the genetic algorithm}, 
  year={2016},
  volume={},
  number={},
  pages={1-7},
}

@article{Lee2001Integration,
  author       = {Lee, H. and Kim, S. S.},
  title        = {Integration of Process Planning and Scheduling Using Simulation Based Genetic Algorithms},
  journal      = {IJAMT},
  volume       = {18},
  number       = {},
  pages        = {586--590},
  year         = {2001},
  publisher    = {Springer},
}

@article{SI201872,
title = {A Petri Nets based Generic Genetic Algorithm framework for resource optimization in business processes},
journal = {Simul Model Pract Theory},
volume = {86},
pages = {72-101},
year = {2018},
issn = {1569-190X},
author = {Yain-Whar Si and Veng-Ian Chan and Marlon Dumas and Defu Zhang},
}

@article{MiddelhuisBSBAD25,
  author       = {Jeroen Middelhuis and
                  Riccardo Lo Bianco and
                  Eliran Sherzer and
                  Zaharah Bukhsh and
                  Ivo Adan and
                  Remco M. Dijkman},
  title        = {Learning policies for resource allocation in business processes},
  journal      = {Inf. Syst.},
  volume       = {128},
  pages        = {102492},
  year         = {2025},
  bibsource    = {dblp computer science bibliography, https://dblp.org}
}

@inproceedings{ZbikowskiOG22,
  author       = {Kamil Zbikowski and
                  Michal Ostapowicz and
                  Piotr Gawrysiak},
  title        = {Deep Reinforcement Learning for Resource Allocation in Business Processes},
  booktitle    = {ICPM},
  novolume       = {468},
  pages        = {177--189},
  year         = {2022},
  bibsource    = {dblp computer science bibliography, https://dblp.org}
}

@article{Kirchdorfer2025,
  author       = {Lukas Kirchdorfer and
                  Robert Blümel and
                  Tomitheus Kampik and
                  Han van der Aa and
                 Heiner Stuckenschmidt},
  title        = {Discovering multi-agent systems for resource-centric business process
                  simulation},
  journal      = {Process Science},
  volume       = {2},
  pages        = {4},
  year         = {2025},
}

@article{SongA08,
  author       = {Minseok Song and
                  Wil M. P. van der Aalst},
  title        = {Towards comprehensive support for organizational mining},
  journal      = {Decis. Support Syst.},
  volume       = {46},
  number       = {1},
  pages        = {300--317},
  year         = {2008},
  bibsource    = {dblp computer science bibliography, https://dblp.org}
}

@INPROCEEDINGS{Peters21,
  author={Peters, S. P. F. and Dijkman, R. M. and Grefen, P. W. P. J.},
  booktitle={EDOC}, 
  title={Resource Optimization in Business Processes}, 
  year={2021},
  volume={},
  number={},
  pages={104-113},
}

@article{Filatov_Yerokhin_2023, 
title={IMPROVED MULTI-OBJECTIVE OPTIMIZATION IN BUSINESS PROCESS MANAGEMENT USING R-NSGA-II }, 
number={3}, 
journal={RADIO ELECTRON COMPU}, 
author={Filatov, V. O. and Yerokhin, M. A.}, 
year={2023}, 
nopages={187} 
}

@ARTICLE{Bejarano23, 
AUTHOR={Bejarano, Jorge  and Barón, Daniel  and González-Rojas, Oscar  and Camargo, Manuel },          
TITLE={Discovering optimal resource allocations for what-if scenarios using data-driven simulation},        
JOURNAL={Frontiers in Computer Science},         
VOLUME={Volume 5}, 
YEAR={2023},
}
\end{document}